\documentclass[12pt]{iopart}
\usepackage{lscape}  
\usepackage{amssymb}  
\usepackage{graphicx}  

\begin{document}
\title[Spectra of W$^{39+}$-W$^{47+}$ in the 12~nm to 20~nm region]
{Spectra of W$^{39+}$-W$^{47+}$ in the 12~nm to 20~nm region observed with
an EBIT light source}

\author{Yu Ralchenko, J Reader, JM Pomeroy, JN Tan, and JD Gillaspy}

\address{National Institute of Standards and Technology, Gaithersburg MD
20899-8422, USA}
\ead{yuri.ralchenko@nist.gov}
\begin{abstract}
We observed spectra of highly ionized tungsten in the extreme ultraviolet 
with an electron beam ion trap (EBIT) and a grazing incidence
spectrometer at the National Institute of Standards and Technology.  
Stages of ionization were distinguished by varying the energy of the electron beam
between 2.1~keV and 4.3~keV and correlating
the energies with spectral line emergence.  The spectra were calibrated by
reference lines of highly ionized iron produced in the EBIT. Identification of
the observed lines was aided by collisional-radiative modeling of the EBIT
plasma. Good quantitative agreement was obtained between the modeling results
and the experimental observations.  Our line identifications complement recent
results for
W$^{40+}$-W$^{45+}$ observed in a tokamak plasma by P\"{u}tterich {\it et al}
(\jpb {\bf 38}, 3071, 2005). 
For most lines we agree with their assignment of ionization stage. Additionally, we
present new identifications for some allowed and forbidden lines of
W$^{39+}$,  W$^{44+}$, W$^{46+}$, and W$^{47+}$.   The uncertainties of our wavelengths 
range from 0.002 nm to 0.010 nm.
\end{abstract} 
\submitto{\JPB} 

\maketitle

\section{Introduction}

Spectra of highly ionized tungsten continue to be of interest for the
development of magnetic fusion energy. Tungsten is considered a strong candidate
for use as a plasma-facing material in advanced tokamaks such as ITER.
Furthermore, accurately measured spectra of highly-charged
ions serve as a test-bed for advanced atomic structure and plasma kinetics
codes.

Spectra of highly-ionized tungsten in the 4~nm to 14~nm region, observed in the
ASDEX Upgrade tokamak in Garching, Germany, were recently reported by
P\"{u}tterich {\it et al} \cite{put}.  This tokamak had part of its first wall
coated with tungsten. In the 12~nm -14~nm region twelve lines of
W$^{40+}$-W$^{45+}$ were identified.  In that measurement, the lines were
observed with a 2.2 m grazing incidence spectrometer.  Ionization stages were
identified by observing the evolution of the lines during the tokamak discharge.
The wavelength uncertainty was given as $\pm$0.005 nm.  These observations made
it possible for the authors of Ref. \cite{put} to resolve a number of questions
that had been raised about line identifications in W spectra from earlier
tokamak work \cite{Ass}. 

In the present work we observed highly ionized W using the electron beam ion
trap (EBIT) operating at the National Institute of Standards and Technology
(NIST). This work is an extension to longer wavelengths of our recent study
\cite{PRA} of the x-ray spectra of W. The present spectra were observed in the
(4-20)~nm region with the grazing incidence spectrometer \cite{Blago}.  As the
(4-8.5)~nm spectrum of tungsten was well covered in a recent report from EBIT-II
at Lawrence Livermore National Laboratory \cite{LLNL}, we concentrate primarily
on the spectra in the longer wavelength region. These measurements complement
the tokamak work \cite{put}  and the new compilation of spectra for all W ions
from W$^{2+}$ to W$^{73+}$ \cite{Kra}. 

\section{Experiment}{\label{exp}}

Highly ionized iron and tungsten were produced in the NIST EBIT.  The EBIT is a
versatile light source, capable of producing nearly any ion charge state of
nearly any element.  The EBIT's  electron beam has a very narrow energy
distribution ($\lesssim$ 60~eV) \cite{beam_width} that allows precise control of
the charge state distribution in the trap.  A detailed description of the NIST
EBIT and its performance is given by one of us elsewhere \cite{EBIT}.  Other
pertinent EBIT parameters in this experiment are a 220 V trap depth, 2.7 T
magnetic flux density, 4 s trap reloading period, and $<$ $2 \times 10^{-3}$ Pa
injected gas reservoir pressure (P1, outer chamber of gas injector, see Ref.
\cite{Fahy} for gas injector description; in the present work, the diameter of
the injector aperture was 3/16 inch, and the inner diameter of the nozzles were
1/8 inch) on a nominal background pressure inside the ion trap of $<$ $5 \times 10^{-9}$ Pa. For
the present experiment, the tungsten atoms were injected into the plasma by a
metal vapor vacuum arc (MEVVA) ion source \cite{Brown}. The MEVVA uses eight
cathodes, any one of which can be quickly selected without disturbing any other
experimental conditions.   This helps to minimize systematic errors in the
calibration. 

The spectra were recorded using a flat-field spectrometer recently deployed on
this facility. It is described thoroughly in a separate report \cite{Blago}. 
Briefly, it consists of a spherical focusing mirror, a bilateral entrance slit,
a gold-coated concave grating, and a liquid-nitrogen-cooled detector. The
focusing mirror has a radius of curvature of 9171~mm and is used at a grazing
angle of $3^\circ$. Together these components maximize the light collection
efficiency while maintaining moderately good spectral resolution. The
diffraction grating, which is the type designed by Harada and Kita
\cite{Harada}, has a radius of curvature of 5649~mm and a variable groove
spacing averaging 1200~lines/mm. When used at a grazing angle of 3$^\circ$, the
reciprocal linear dispersion at 12~nm is about 0.6~nm/mm.  The detector is a
back-illuminated charge-coupled device (CCD) camera placed at the grating's
focal surface.  The CCD consists of a 1340$\times$400 array of  pixels that are
directly exposed to extreme ultraviolet (EUV) radiation. The signals are 
column-summed in hardware to minimize read-out noise.  Observations of the full
frame exhibit negligible spectral line curvature.  The instrument's resolution
as configured for this experiment is about 350, corresponding to a resolving
limit of about 0.03 nm. 

\begin{figure}
\includegraphics{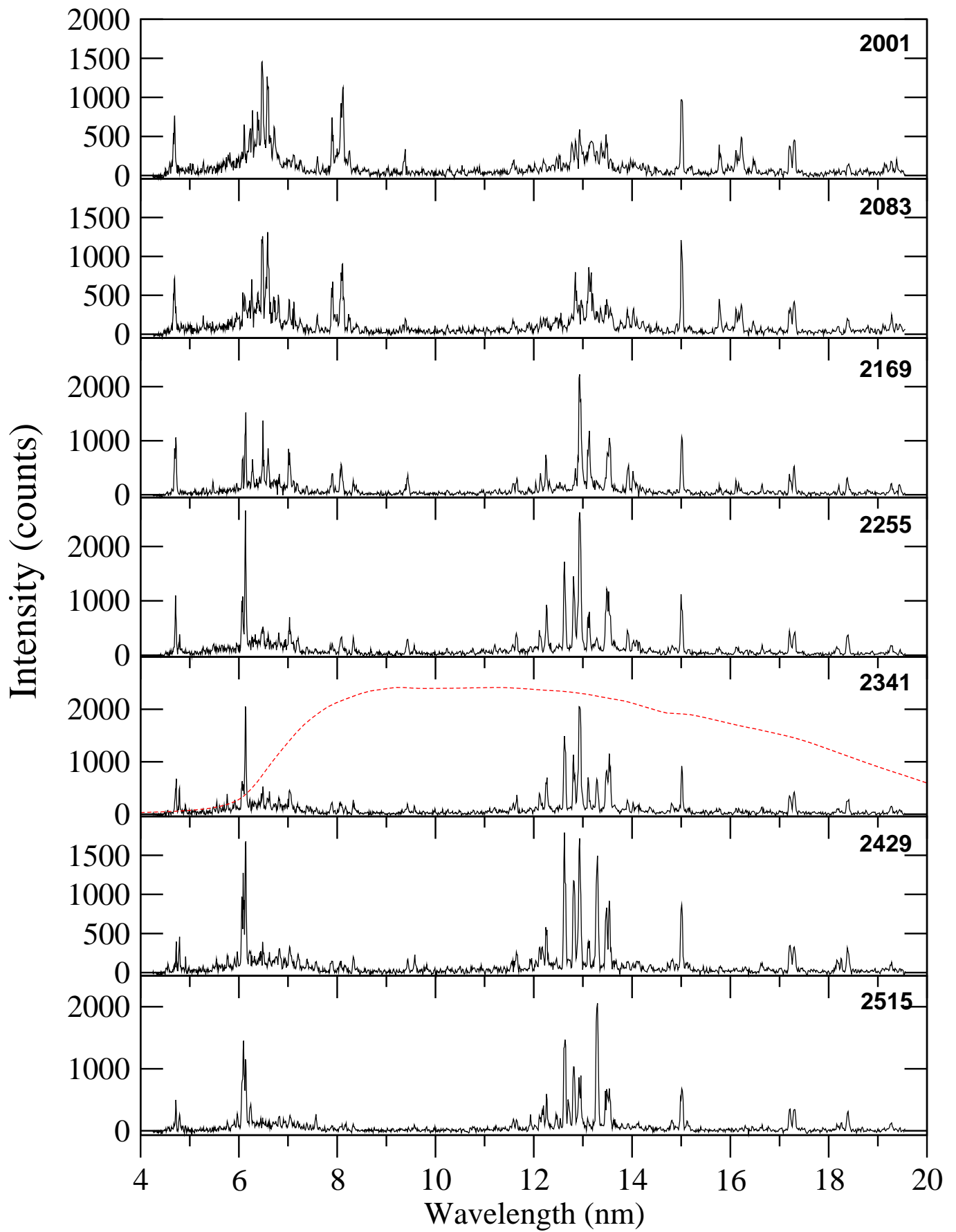}
\caption{Experimental spectra for low beam energies of run A. The dashed line
represents a relative efficiency curve for the spectrometer.}
\label{fig:fig1}
\end{figure}

Light from wavelengths above 25 nm was filtered out by a zirconium foil  placed
between the EBIT and the grazing-incidence focusing mirror. An efficiency curve
for the  entire spectral recording system (including filter) is shown by the
dashed line in Fig. \ref{fig:fig1}. As seen, the system has good transmission
between 6 and 18 nm. The maximum transmission is about 60 \% at 10 nm
\cite{Blago}. Even without the Zr filter, as shown in Ref. \cite{Blago}, the
detection efficiency of the spectrometer falls off rapidly below 8 nm.

The tungsten spectra were calibrated by lines of  Fe$^{17+}$-Fe$^{22+}$, with
wavelengths taken from the NIST Atomic Spectra Database (ASD) \cite{ASD}. For
the calibration spectra, iron was injected into the EBIT plasma periodically
during the course of experimental runs. The iron ions were excited at several
beam energies between 1.8~keV and 2.9~keV. We also used spectral lines of
O$^{4+}$ and O$^{5+}$, with wavelengths taken from ASD, as well as second order
lines of W in the 6~nm to 8~nm region, with wavelengths taken from Ref.
\cite{LLNL}. The oxygen was likely introduced into the plasma through a side
port, to which another spectrometer was connected.

In order to distinguish stages of ionization, we took a series of spectra at
nineteen different beam energies between 2.0~keV and 4.3~keV. The emergence of
new spectral lines as the energy was changed was correlated with the ionization
energies of the tungsten ions. The ionization energies of the ions relevant to
this work as determined by Kramida and Reader \cite{KR} are shown in Table
\ref{Table1}.  Our ionization stage assignments were also correlated with the
stages of lines in the 4~nm to 8~nm region given in Ref. \cite{LLNL}, where a
similar method was used. 

\begin{table}
\caption{\label{Table1}Ionization potentials (in eV) of W ions.}
\begin{indented}
\lineup
\item[]\begin{tabular}{@{}lcllc}
\br
Ion & Sequence & Ground state & Ref. \cite{KR} & Present work\\
 &  & configuration & & (calculated)\\
\mr
W$^{40+}$ & Se & $4s^24p^4$ 	& 1940.6 $\pm$ 2.0 & 1944 \\
W$^{41+}$ & As & $4s^24p^3$ 	& 1994.8 $\pm$ 2.0 & 1997 \\
W$^{42+}$ & Ge & $4s^24p^2$ 	& 2149.2 $\pm$ 2.1 & 2147 \\
W$^{43+}$ & Ga & $4s^24p$ 		& 2210.0 $\pm$ 1.5 & 2206 \\
W$^{44+}$ & Zn & $3d^{10}4s^2$ 	& 2354.5 $\pm$ 1.4 & 2354 \\
W$^{45+}$ & Cu & $3d^{10}4s$ 	& 2414.1 $\pm$ 0.4 & 2415 \\
W$^{46+}$ & Ni & $3d^{10}$ 		& 4057 $\pm$ 3   & 4052   \\
W$^{47+}$ & Co & $3d^{9}$ 		& 4180 $\pm$ 4   & 4173   \\
W$^{48+}$ & Fe & $3d^{8}$ 		& 4309 $\pm$ 4   & 4304   \\
\br
\end{tabular}
\end{indented}
\end{table}

\begin{figure}
\includegraphics{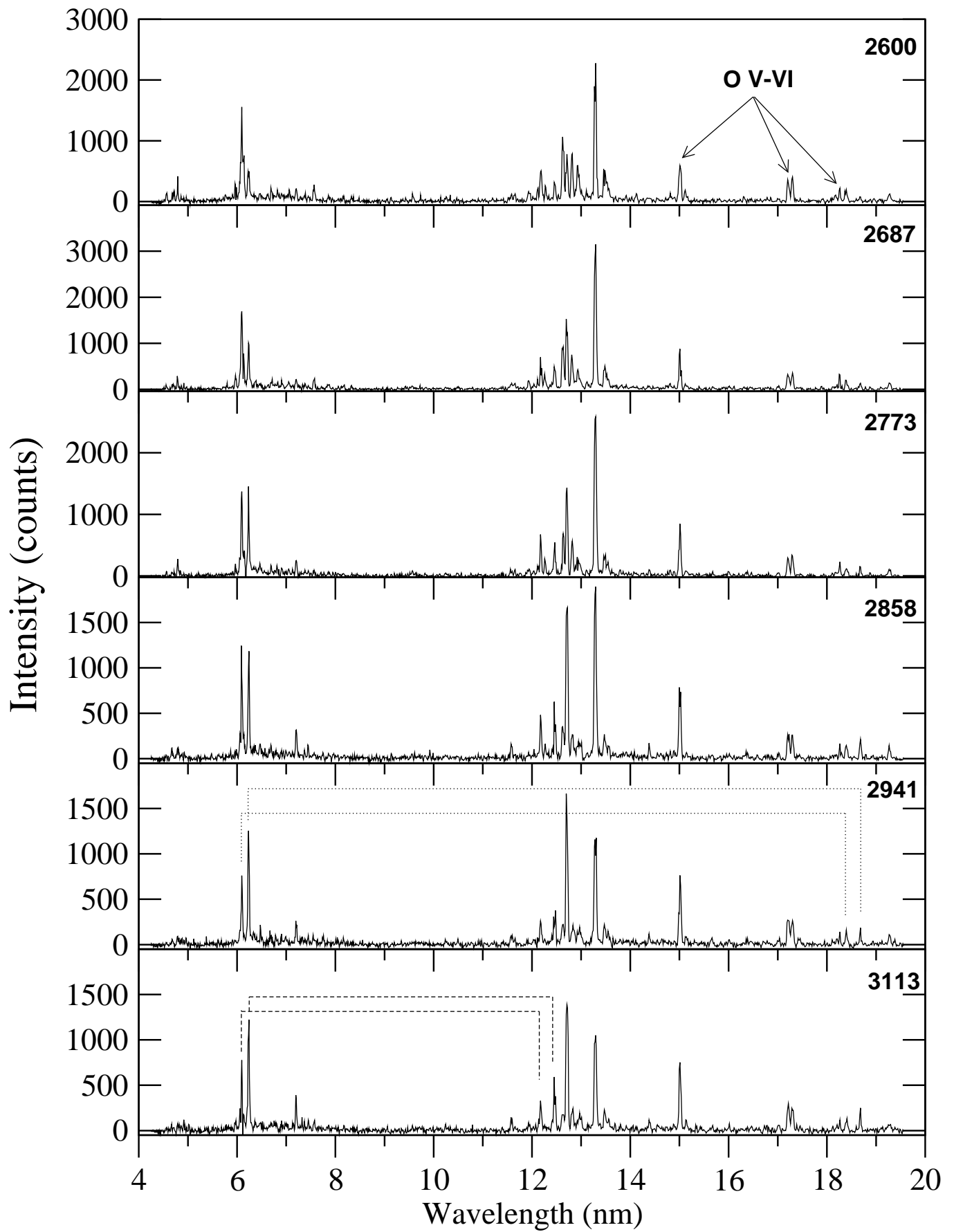}
\caption{Experimental spectra for high beam energies of run A. Examples of higher
order spectral lines are indicated by dashed (second-order) and dotted (third-order)
lines.}
\label{fig:fig2}
\end{figure}

\begin{figure}
\includegraphics{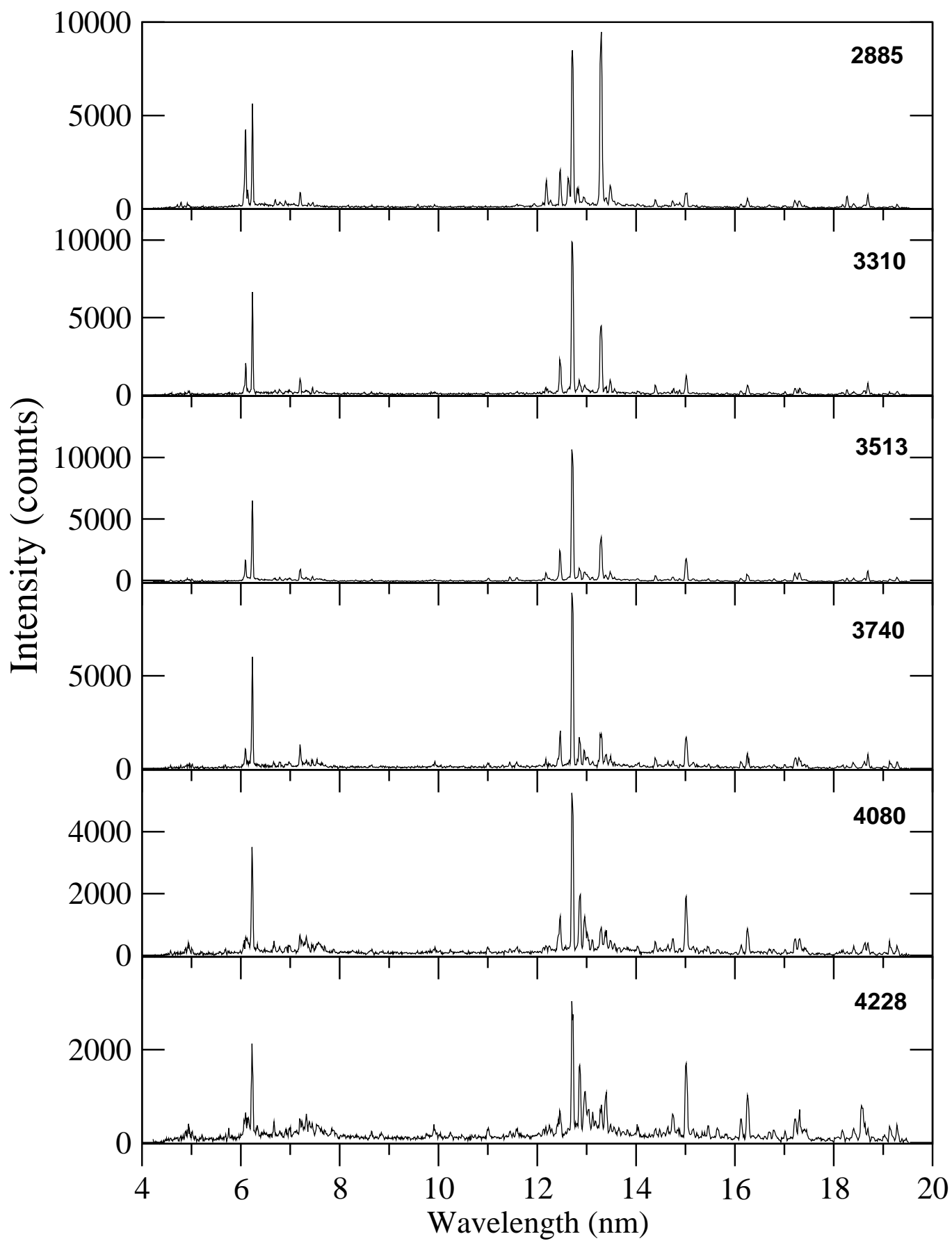}
\caption{Experimental spectra for run B.}
\label{fig:fig3}
\end{figure}

The spectra were measured during two separate runs. The first one (run A)
covered 13 beam energies from 2001~eV to 3113~eV. Spectra from this run are
shown in Figs. \ref{fig:fig1} and \ref{fig:fig2} after a background (dominated
by essentially noiseless CCD bias) of approximately 4000 counts/channel was
subtracted. The second run (run B) covered 6 energies from 2885~eV to 4228~eV,
as shown in Fig. \ref{fig:fig3}. The nominal beam energies and currents are
presented in Table \ref{TableR}. In what follows, the spectra will be referred
to by the beam energy followed by the run identification, e.g., 2001A.  For
completeness and easier identification of the higher-order lines, the figures
show the whole spectral range from 4 nm to 20 nm.

\begin{table}
\caption{\label{TableR}Nominal beam energies (in eV) and currents (in mA) for
runs A and B.}
\begin{indented}
\lineup
\item[]\begin{tabular}{@{}rrrrrr}
\br
\multicolumn{4}{c}{Run A} & \multicolumn{2}{c}{Run B}\\
\crule{4} & \crule{2}\\
\\
Energy & Current & Energy & Current & Energy & Current\\
\mr
2001 & 28.9 & 2600 & 32.9 & 2885 & 52.0  \\
2083 & 31.5 & 2687 & 32.9 & 3310 & 67.0  \\
2169 & 32.5 & 2773 & 32.6 & 3513 & 78.0  \\
2255 & 32.5 & 2858 & 32.6 & 3740 & 85.5  \\
2341 & 32.7 & 2941 & 32.7 & 4080 & 94.5  \\
2429 & 32.8 & 3113 & 32.7 & 4228 & 103.0 \\
2515 & 32.9 &      &      &      &       \\
\mr
\end{tabular}
\end{indented}
\end{table}

\section{Calculations of spectra}

Similar to our work on x-ray emission \cite{PRA}, the analysis of the EUV
spectra was aided by collisional-radiative (CR) modeling of the EBIT plasma. The
non-Maxwellian CR code {\sc NOMAD} \cite{NOMAD} was used to calculate the level
populations and line intensities for various electron beam energies. The
electron density was fixed in the model at N$_e$ = 10$^{11}$ cm$^{-3}$. Test
runs showed that the modeling results are not sensitive to the value of N$_e$ 
in the range 10$^{11}$ cm$^{-3}$ to 10$^{12}$ cm$^{-3}$.  

Since the atomic data required for the CR modeling of highly-charged tungsten
are not available in the literature, we performed large-scale calculations of
all relevant parameters, i.e., energy levels, radiative transition probabilities
(including forbidden transitions), and collisional cross sections, using the
relativistic Flexible Atomic Code ({\sc FAC}). The {\sc FAC} methods and
techniques, which are described in detail in Ref. \cite{FAC}, are highly
suitable for multiply-charged ions of heavy elements, clearly of great
importance for the present work. Although the calculations are carried out in a
relativistic jj-coupling scheme, we will also be using LS-coupling notation if
this is more appropriate. As the CR modeling of non-Maxwellian plasmas, such as
that of EBIT, is based on collisional cross sections rather than
thermally-averaged rate coefficients, we created a database of cross sections
for electron impact excitation, ionization, and radiative recombination. The
cross sections of inverse processes were determined from the principle of
detailed balance. Dielectronic recombination was not important in this
experiment because the narrow  electron-energy distribution function (EEDF) of
the beam does not overlap with the resonant energies that are of importance for
dielectronic capture. Also, three-body recombination can be completely neglected
under the present low-density conditions.

A total of about 5000 atomic levels for ionization stages from Br-like W$^{39+}$
to Mn-like W$^{49+}$ were included in our simulations. The EEDF of the electron
beam was modeled by a Gaussian function with a full width at half-maximum of 60
eV. The results of the calculations were not sensitive to the beam energy width
in the range of 30 eV to 60 eV, a range that covers that typical of an EBIT.

\subsection{Atomic data accuracy}

The assessment of the accuracy of the atomic data for highly-ionized tungsten is
impeded by the lack of reliable experimental information, especially for
collisional cross sections. Nonetheless, good agreement between the experimental
and simulated x-ray spectra \cite{PRA} provides a strong indication that the 
data generated by {\sc FAC} are sufficiently accurate for our present purpose.
One may generally expect higher precision for x-ray spectra simulations compared
to the EUV case due to higher importance of correlation effects for EUV
transitions between excited states. Nevertheless, as will be shown below, the
simulated EUV spectra are also in good agreement with the measurements, both
for line positions and for intensities. Another test of the {\sc FAC} data is
provided by the calculated ionization potentials (last column in Table
\ref{Table1}), which are seen to differ from the recommended values of Ref.
\cite{KR} by less than 0.2~\%.

Although several papers calculating energy levels and radiative transition
probabilities for [Ga] and [Ni] isoelectronic sequences were published recently
(e.g., \cite{Quinet07,Saf06,Dong}), the most comprehensive large-scale
calculation of atomic data and spectral line intensities for highly-ionized
tungsten remains that by Fournier \cite{Fournier}. He used the fully
relativistic parametric potential code {\sc RELAC} \cite{RELAC1,RELAC2,RELAC3}
to produce spectroscopic data for ions of tungsten from Rb-like W$^{37+}$ to
Co-like $W^{47+}$.

\subsection{Collisional-radiative modeling}

The modeling was performed in the steady-state equilibrium approximation.  In
order to test the applicability of this approximation, we used {\sc NOMAD}'s
ability to solve time-dependent rate equations. It was found that at an electron
density of N$_e$ = 10$^{11}$ cm$^{-3}$, a steady state for a W plasma is reached
within few tenths of a second. As the trap is reloaded typically every 4 s and
observed continuously over the course of this time, our use of the steady-state
approach is fully justified.

The calculated line intensities were convolved with a Gaussian function
representing instrumental resolution and then multiplied by the transmission
function. Figure \ref{fig:fig1} shows that below 7~nm the calculated
transmission function drops rapidly, and in this region it is known less
accurately than for longer wavelengths. We believe this uncertainty explains the
somewhat larger discrepancies between intensities of the experimental and
simulated spectra for shorter wavelengths.

The theoretical beam energy, used as a free parameter, was varied until a good
fit of calculated and measured spectra was obtained. In all simulations the
fitted theoretical energy was found to be smaller than the nominal beam energy.
For instance, the energy difference for run A increases from several tens of~eV
at the lowest beam energies to about 500~eV at E = 3113 eV. Such a difference is
likely due to (i) space charge effects  and (ii) charge exchange (CX) with
neutrals or low-charged ions that are present in the EBIT. The space charge
effect is estimated to reduce the energy of the beam electrons by not more than
about 150 eV. While CX does not modify the beam energy, it effectively enhances
recombination and thus shifts ionization balance toward lower charge states. To
test the importance of CX, we performed a series of {\sc NOMAD} runs using a
simple approximation for the CX cross sections \cite{Elton} and varying both the
relative velocity of neutrals and W ions, $v_{rel}$, and the density of neutrals
$N_n$. Although this modification significantly improved the agreement between
experimental and fitted beam energies, the uncertainties in $v_{rel}$ and $N_n$
leave too many free parameters to unambiguously account for the effect of charge
exchange. Improved models of EBIT trap dynamics are needed to determine these
parameters. Hence, we excluded the CX contribution in our simulations. 

Although CX affects the ionization balance, it is not likely to modify the
relative line intensities within individual ion stages. The CX of neutrals with
highly-charged ions results in preferential population of the shell with the
principal quantum number $n \approx Z_{c}^{3/4}$ \cite{Janev} ($Z_c$ is the ion
charge), which gives $n \approx 16-18$ for $Z_c$ = 40-47. The ensuing radiative
cascades are likely to smear out the population flux such that the {\it
relative} populations of the low-excited states, which are responsible for the
EUV emission, would not be modified. An indirect confirmation of this conclusion
follows from our previous work on x-ray spectra \cite{PRA} where the relative
intensities of the W$^{46+}$lines originating from the $n=4-6$ shells were
accurately calculated without involving charge exchange effects. 

Finally, note that the variation of the electron beam energy by a few hundred eV
does not noticeably affect the relative distribution of line intensities
within a specific ion. Therefore, one could separately calculate spectral emission
patterns for each ion and then derive the experimental ionization distribution
using ratios of line intensities originating from different ionization stages.

\section{Experimental results and interpretation}{\label{results}}

All identified lines in the measured spectra of W (Figs.
\ref{fig:fig1}-\ref{fig:fig3}) represent $\Delta n =0$ transitions with 
principal quantum number $n=4$. The low electron density of the EBIT plasma
permits measurement of both allowed and forbidden lines. Strong impurity lines
from the highly-ionized ions of oxygen and nitrogen are easily recognized by
their almost constant intensity as the beam energy is changed. For instance, the
lines at 15.01 nm, 17.21 nm, and 17.30~nm noted in the 2600A spectrum in Fig.
\ref{fig:fig2}, are the well known lines of O V and O VI \cite{ASD}. The
nitrogen was deliberately introduced as an ion-cooling agent.  Also, the
strong lines near 6 nm can be seen in the second and third grating orders. The
high-order lines are indicated by dashed (second order) and dotted (third order)
lines in Fig. \ref{fig:fig2}.

The list of identified lines with measured wavelengths is given in Table
\ref{Table2}. This table also includes other experimental \cite{put} and
calculated \cite{Fournier} wavelengths as well as the weighted oscillator
strengths $gf$. The agreement between $gf$ values calculated by {\sc RELAC}
\cite{Fournier} and {\sc FAC} is excellent for almost all reported lines. The
only exception is the magnetic-dipole (M1) line $3d^9$
$^{2}D_{3/2}$--$^{2}D_{1/2}$ in the Co-like ion, where the discrepancy is about 
20~\%.

\subsection{Low-energy spectra: 2000 eV to 2600 eV}

As seen from Table \ref{Table1}, the lowest beam energies of 2001~eV and 2083~eV
are sufficient to produce ions up to Ge-like W$^{42+}$, with ground state
configurations $4s^24p^k$, $k \geq 2$. These ions have a relatively large number
of $n=4$ excited states, which results in appearance of wide transition arrays
near 6~nm and 13~nm, rather than well separated spectral lines. From comparison
with the calculated spectra it was, nonetheless, possible to identify a new
magnetic-dipole line $4p^5~^2P_{3/2} - 4p^5~^2P_{1/2}$ in Br-like
W$^{+39}$.

\begin{figure}
\includegraphics[width=\textwidth]{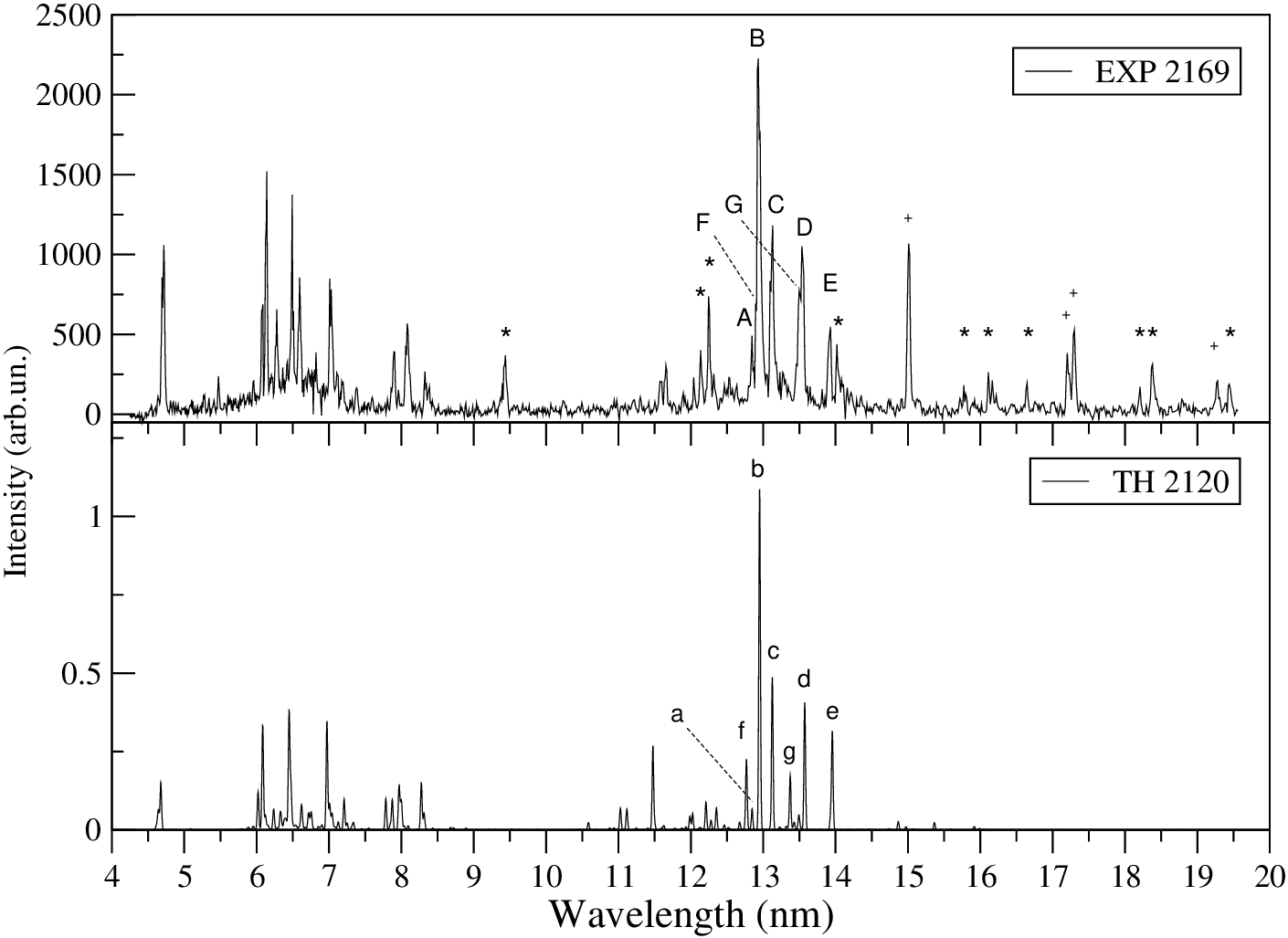}
\caption{Comparison of the experimental (top) and calculated (bottom) spectra at
the nominal beam energy of 2169 eV. The high-order lines are marked by stars,
and the impurity lines are marked by crosses.}
\label{fig:fig4}
\end{figure}

A dramatic change in the appearance of the spectrum is seen in Fig.
\ref{fig:fig1} when the beam energy is changed from 2083~eV to 2255~eV. At the
higher energy, which is sufficient to reach all ionization stages up to Zn-like
W$^{44+}$ (yet not to produce a significant amount of this ion), individual
spectral lines become more visible in the spectrum. The agreement between
experimental and simulated spectra patterns is exemplified in Fig.
\ref{fig:fig4}. While the nominal beam energy was 2169~eV, the best theoretical
fit, based on the relative intensities of  lines from different W ions in the
12~nm to 14~nm region, was obtained for an energy of 2120~eV. Aside from
impurity and second-order lines that are marked in Fig. \ref{fig:fig4} by
crosses and stars, respectively, almost all measured lines can be matched to
calculated ones from a visual comparison of line positions and their
relative intensities. Moreover, the derived line identifications are well
confirmed by the dependence of line intensities upon the beam energy. 

As an example, consider in more detail the identification procedure for the
strong lines between 12~nm and 14~nm shown in Fig. \ref{fig:fig4} (see also
Table \ref{Table2}). Our collisional-radiative modeling shows that at the fitted
energy of 2120~eV the most populated ions of tungsten have the following
distribution: [Br] : [Se] : [As] : [Ge] = 0.01 : 0.10 : 0.42 : 0.46, and thus
one may expect to observe strong emission in W$^{41+}$ and W$^{42+}$. The
strongest lines are marked by letters A through E in Fig. \ref{fig:fig4}. Both
the positions and relative intensities of the lines B, C, D, and E very well
agree with the calculated ones ($b$, $c$, $d$, $e$), and therefore these four
lines can be identified as follows: B - electric-quadrupole (E2) $4p^{2~3}P_0 -
4p^{2~1}D_2$ transition in Ge-like W$^{42+}$, C - M1 $4p^{3~2}D_{3/2} -
4p^{3~2}D_{5/2}$ transition in As-like W$^{41+}$, D - M1 $4p^{2~3}P_0 -
4p^{2~3}P_1$ transition in W$^{42+}$, E - M1 $4p^{3~2}D_{3/2} - 4p^{3~4}S_{3/2}$
transition in W$^{41+}$.  This identification is strengthened by the fact that,
as seen from Fig. \ref{fig:fig1}, lines C and E become relatively weaker than
lines B and D as the beam energy increases. In Ref. \cite{put} line D was
associated with the Ga-like ion.

The assignment of quantum numbers for the W$^{41+}$ levels is highly
problematic due to very strong mixing of the basis states in the ground
configuration. For instance, although the ground level is designated
$4s^{2}4p^{3~2}D_{3/2}$, it actually has only 26~\% $^{2}D$ character
\cite{Kra}.

\begin{figure}
\centering
\includegraphics[width=0.6\textwidth]{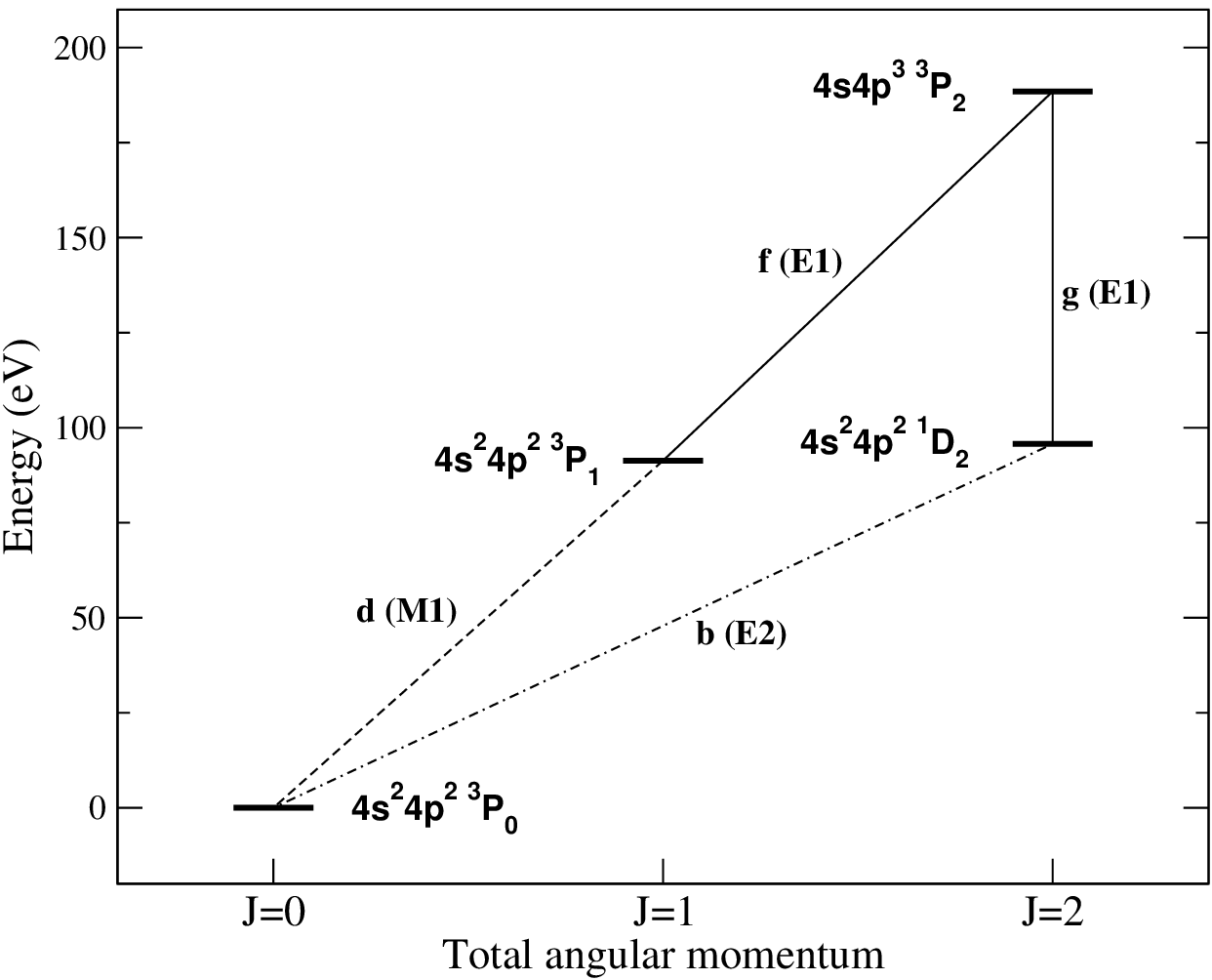}{}
\caption{Four lowest energy levels in Ge-like W$^{42+}$. The line labels (see
Fig. \ref{fig:fig4}) and transition
multipoles are indicated next to the lines.}
\label{fig:fig_en}
\end{figure}

The theoretical spectrum in Fig. \ref{fig:fig4} shows the presence of two other
lines, {\textit f} and {\textit g}, which correspond to transitions $4p^{2~3}P_1
- 4s4p^3~^{3}P_2$ and $4p^{2~1}D_2 - 4s4p^3~^{3}P_2$, in the Ge-like
ion\footnote{The upper level, which in fact is the lowest odd level with $J=2$,
is identified as $4s(^{2}S_{1/2})4p^{3}(^{2}P^{\circ}_{3/2}) ~
(1/2,3/2)^{\circ}_{2}$ in Ref. \cite{Kra}.}. Although it may appear that the
experimental line A should be associated with {\textit f}, this line can also be
seen at lower energy  and therefore is assigned to the transition $4p^{4~3}P_2 -
4p^{4~1}D_2$ in Se-like W$^{40+}$.  In order to find the experimental lines
corresponding to  {\textit f} and {\textit g}, one can make use of the following
observation. The {\it four} strongest lines from W$^{42+}$ near 13 nm, which are
listed in Table \ref{Table2}, connect {\it four} levels (Fig. \ref{fig:fig_en}).
Therefore, if one denotes the transition energy corresponding to line {\textit
a} as $E(a)$, then $E(f)+E(d)$ = $E(b)+E(g)$, or $E(b)-E(d)$ = $E(f)-E(g)$.
Hence, as $E(b)-E(d)$ can be easily determined from the experimental spectra,
the other pair of lines can be found using the derived difference of
wavelengths. While no isolated spectral lines near 13 nm seem to satisfy this
condition, two blended structures on the short-wavelength shoulders of lines B
and E indeed correspond to the derived separation. Therefore we tentatively 
identify this pair of lines, which are shown by the upper-case letters F and G
in Fig. \ref{fig:fig4}, with the transitions {\textit f} and {\textit
g}.

\begin{figure}
\centering
\includegraphics[width=0.6\textwidth]{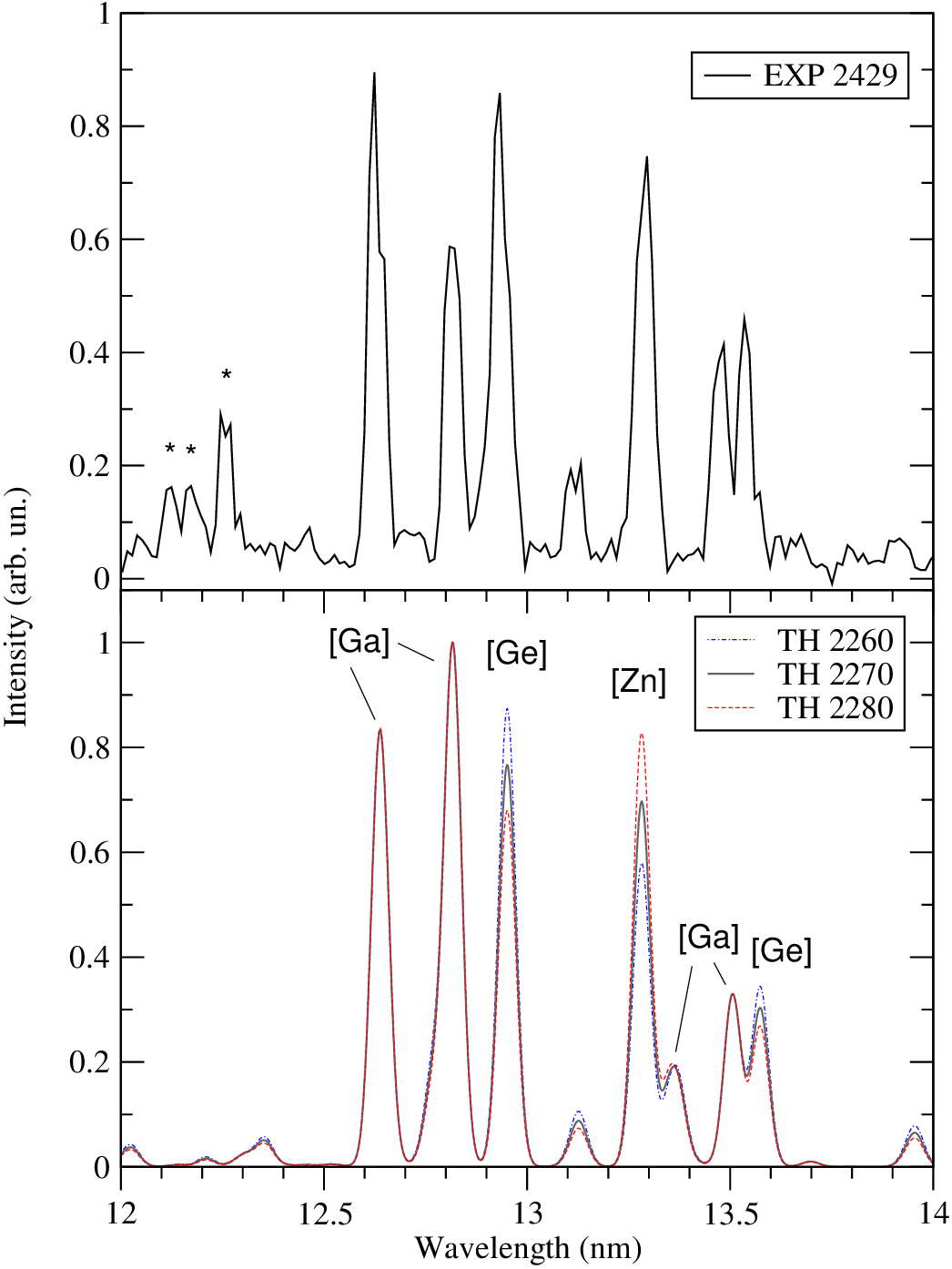}
\caption{Comparison of the experimental (top) and calculated (bottom) spectra at
the nominal beam energy of 2429 eV. The second-order lines are marked by stars.}
\label{fig:fig2429}
\end{figure}

As the beam energy varies between 2000 eV and 2600 eV, new ionization stages of
W appear in the spectrum and the spectral patterns change markedly. Figure
\ref{fig:fig2429} (top) shows the measured spectrum at 2429 eV from 12 nm to
14 nm. The best fit was obtained for a fitted beam energy of 2270 eV (bottom of
Fig. \ref{fig:fig2429}), and one can see a good correspondence between the
simulations and the measured spectrum. At this energy, the lines from the Ge-
and Ga-like ions are the strongest, and their identification poses no difficulty
(see Table \ref{Table2}). Since the beam energy exceeds the ionization potential
of the Ga-like ion, the intercombination $4s^{2}~^{1}S_{0} - 4s4p~^{3}P_{1}$
line in Zn-like W$^{44+}$ is also visible at 13.288 nm. One can see from this
figure that the relative line intensities are calculated accurately for almost
all lines. The only exception is the  $4s^{2}4p~^{2}P_{1/2} -
4s4p^{2}~^{4}P_{1/2}$ transition in the Ga-like ion at 12.817 nm, for which the
calculated intensity is too high by about a factor of two. Such a discrepancy
was found for all other beam energies where this line is visible.

To illustrate the sensitivity of the simulated spectra to the beam energy, in
Fig. \ref{fig:fig2429} we also give results of the simulations at 2260 eV, 2270
eV, and 2280 eV, normalized to the peak of the 12.817 nm line in the Ga-like
ion. A change of only 10 eV in the theoretical beam energy does not affect the
intensities of the spectral lines within the same ion (cf. Ga-like lines at
12.629 nm, 12.817 nm, and 13.481 nm), but does drastically modify the relative
intensities for different ions. This is most clearly seen when comparing the
[Ge] 12.936 nm and the [Zn] 13.288 nm lines: the intensity ratio changes from
1.50 to 0.82 when the theoretical beam energy increases from 2260 eV to 2280 eV. 

Although the line at 13.481 nm was identified as the [Ge] $4p^{2}~^{3}P_{0} -
4p^{2}~^{3}P_{1}$ transition in Ref. \cite{put}, it is evident both from this
plot and from the variation of line intensity with the beam energy (Fig.
\ref{fig:fig1}) that it originates from the Ga-like ion. Accordingly, we
identify it with the $4s^{2}4p~^{2}P_{3/2} - 4s4p^{2}~^{4}P_{5/2}$ transition.

\subsection{Medium-energy spectra: 2600 eV to 3500 eV}

The spectra for the beam energies between 2600 eV and 3500 eV are dominated by a
strong emission in the $4s^2-4s4p$ and $4s-4p$ lines from Zn- and Cu-like ions,
respectively. The large difference between the ionization potentials of the Cu-
and Ni-like ions is responsible for small modifications in ionization balance,
and thus small variations in the spectral pattern. Although the energy of the
beam is sufficient to produce Ni-like W$^{46+}$, the corresponding $n=4-4$ 
transitions are very weak as the total population of W$^{46+}$ is still
relatively small. 

\begin{figure}
\includegraphics[width=\textwidth]{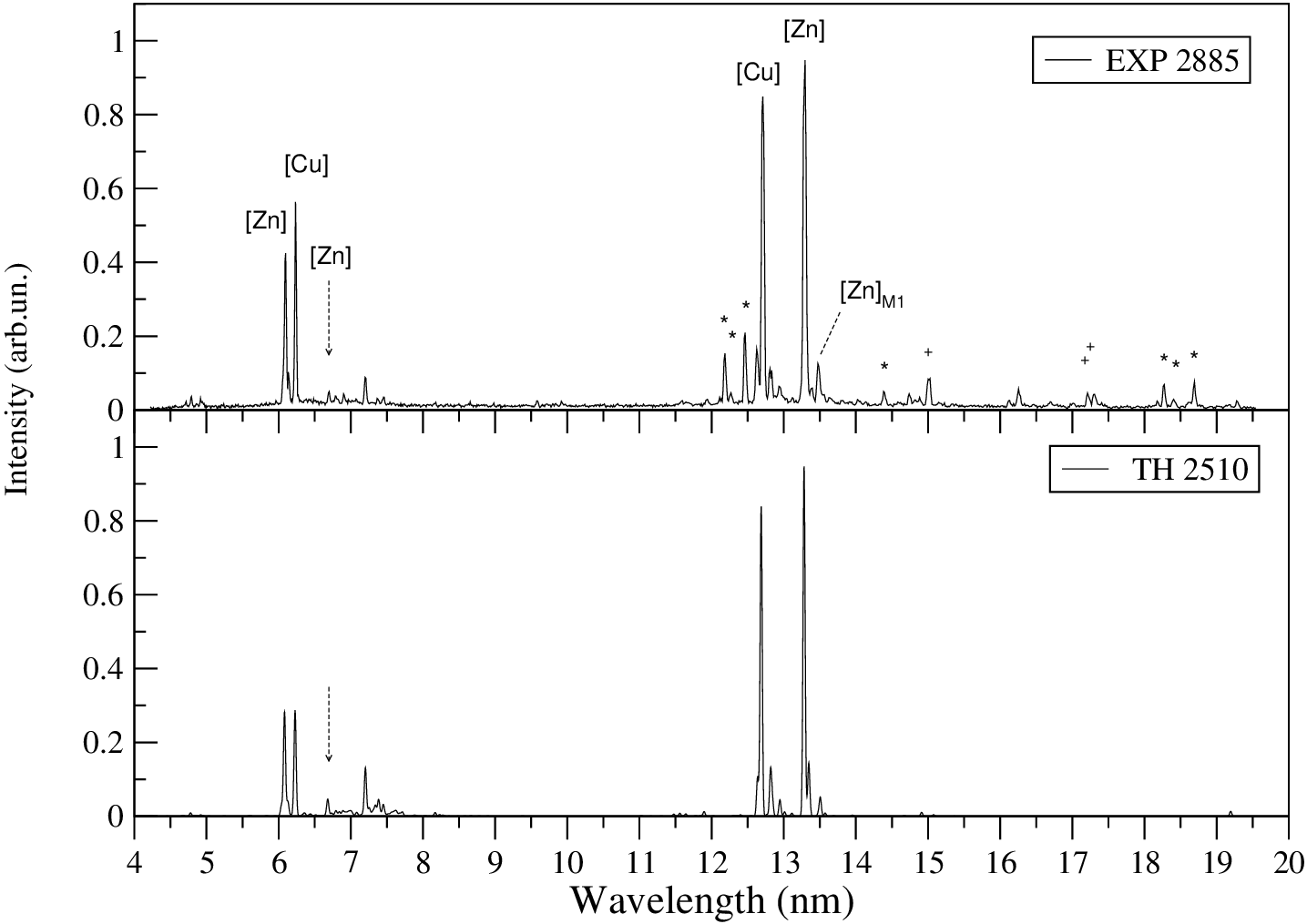}
\caption{Comparison of the experimental (top) and calculated (bottom) spectra at
nominal beam energy of 2885 eV. High-order lines are marked by stars,
and impurity lines are marked by crosses. Arrows indicate the $4s4p~(1/2,1/2)_2 -
4s4d~(1/2,5/2)_2$ line in the Zn-like ion at 6.69301 nm (see text).}
\label{fig:fig2885}
\end{figure}

Figure \ref{fig:fig2885} compares the experimental spectrum at 2885 eV with the
simulated spectrum at 2510 eV. As already mentioned, we believe that this
difference of 375 eV is due to CX and space-charge effects that are difficult to
accurately account for. Other than that, the agreement in line positions and
intensities is very good over the whole range. The well known intercombination
$4s^{2~1}S_{0} - 4s4p~^{3}P_{1}$ line (or $4s^{2}~(1/2,1/2)_{0} -
4s4p~(1/2,1/2)_{1}$ in jj-coupling) from the Zn-like ion and resonance $4s_{1/2}
- 4p_{1/2}$ line from the Cu-like ion are  the most prominent. The other
measured lines between 12.5 nm and 13.5 nm can also be unambiguously identified,
including the magnetic-dipole $4s4p~^3P_1 - 4s4p~^3P_2$ (or $4s4p~(1/2,1/2)_{1}
- 4s4p~(1/2,3/2)_{2}$) line in [Zn] W$^{44+}$ at 13.480 nm. The relative
intensity of this line, which hereafter will be referred to as [Zn]$_{M1}$,
remains approximately constant with respect to the intercombination line at
13.288 nm over a large range of energies indicating that it indeed originates
from the Zn-like ion.  This line was not observed in the tokamak spectrum
\cite{put}. 

Our identification of the [Zn]$_{M1}$ line presents a problem with regard to
some of the identifications of Zn-like lines identified in the LLNL EBIT
measurements \cite{LLNL}.  In Ref. \cite{LLNL}, two Zn-like lines were
identified as $4s4p~(1/2,3/2)_2 - 4s4d~(1/2,5/2)_2$ at 4.45299(62) nm and
$4s4p~(1/2,1/2)_2 - 4s4d~(1/2,5/2)_2$ at 6.69301(40) nm.  This implies a
wavelength for [Zn]$_{M1}$ of 13.305(6) nm, compared to our observed wavelength
of 13.480(3) nm, a clear inconsistency. In Ref. \cite{LLNL} the 4.45299  nm line
was observed with a relative  intensity of 0.04, the same as that of the 6.69301
nm line.  However, both our present calculations and those of Ref.
\cite{Fournier} indicate that the 6.69301 nm line should be stronger by a factor
of nearly 4. Since the upper level is the same for both lines, the calculated
ratio is not sensitive to beam energy or other collisional effects.  Since we
observe the 6.69301 nm line  (indicated by arrows in Fig. \ref{fig:fig2885}) as
was observed in \cite{LLNL}, we suspect that the problem rests with the 4.45299
nm line.  (We are not able to observe this line due to the low transmission of
our filter at this wavelength.)  If we accept the identification of the 6.69301
nm line, our wavelength for [Zn]$_{M1}$ implies a wavelength for
$4s4p~(1/2,3/2)_2 - 4s4d~(1/2,5/2)_2$ of 4.4724(4) nm, and it is likely that the
identification in Ref. \cite{LLNL} will have to be revised.

\subsection{High-energy spectra: 3500 eV to 4228 eV}

For beam energies greater than 3500 eV (Fig. \ref{fig:fig3}), lines from Ni-like
W$^{46+}$ become more intense due to a larger relative population of this ion
and also cascades from highly-excited levels that are populated by electron
excitation. The lines from the lower (Z$_c$ $<$ 44) charge states are barely
visible, while the resonance $4s-4p$ line in the Cu-like ion is still the
strongest.

\begin{figure}
\includegraphics[width=\textwidth]{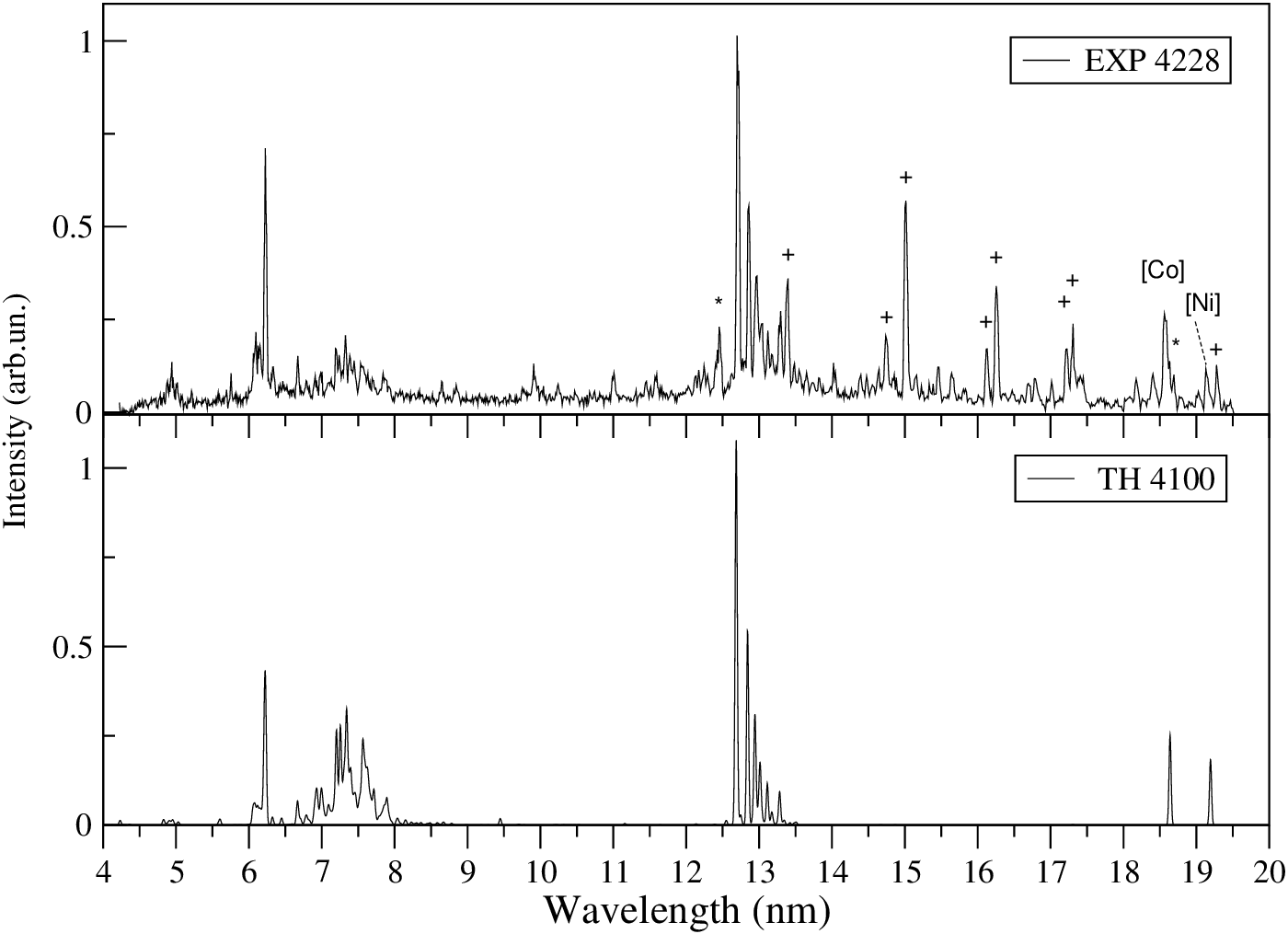}
\caption{Comparison of the experimental (top) and calculated (bottom) spectra at
the nominal beam energy of 4228 eV. The high-order lines are marked by stars,
and the impurity lines are marked by crosses.}
\label{fig:abs}
\end{figure}

The best fit for the measured spectrum at 4228~eV was obtained at a theoretical
beam energy of 4100~eV (see Fig. \ref{fig:abs}). This energy is above the
ionization threshold of the Ni-like W$^{46+}$ and thus one may expect to find
some lines from the Co-like ion W$^{47+}$. The strongest calculated line in
W$^{47+}$  corresponds to the M1 transition $3d^{9}$ $^{2}D_{5/2}$ -- $3d^9$
$^{2}D_{3/2}$ with theoretical wavelength of 18.640~nm. Indeed, the measured
spectrum contains a strong line at 18.578$\pm$0.002~nm, which is absent at lower
energies. We therefore identify this as the [Co] M1 transition.  Twenty years
ago Ekberg et al. \cite{Ekberg}, using observed wavelengths for the
$3p^{6}3d^{9}~^{2}D_{5/2} - 3p^{5}3d^{10}~^{2}P_{3/2}$ and
$3p^{6}3d^{9}~^{2}D_{3/2} - 3p^{5}3d^{10}~^{2}P_{3/2}$ transitions in
lower-charged ions of the isoelectronic sequence, predicted a wavelength for the
$^{2}D_{5/2}$ -- $^{2}D_{3/2}$ transition in W$^{47+}$ to be
18.541$\pm$0.032~nm. The weaker structure at the red wing of this line is the
[Cu] 6.234~nm line \cite{LLNL} in the third order. 

The $n=4-4$ lines of the Ni-like ion are the most prominent in the 4228 eV
spectrum. The wide transition array between 6.5~nm and 8~nm is mainly due to the
$3d^94p - 3d^94d$ and $3d^94d - 3d^94f$ transitions in this ion, although it
also includes tens of overlapping lines from the Cu-like ($4p-4d$) and Co-like
($3d^84p-3d^84d$ and $3d^84d-3d^84f$) ions. Although a good correspondence
between the experimental and calculated spectral features can be seen, 
identification of individual lines is not possible with our spectral resolution.

The well-resolved lines near the [Cu] $4s_{1/2} - 4p_{1/2}$ line at 12.7118~nm,
listed in Table \ref{Table2}, are mainly due to the $3d^94s - 3d^94p$
transitions in W$^{46+}$. According to our calculations, there are only six
strong $3d^94s - 3d^94p$ lines between 12.5 nm and 13.5 nm. The (5/2,1/2)$_3$ --
(5/2,1/2)$_3$ transition is completely blended with the [Cu] 12.7118-nm line,
while the (3/2,1/2)$_1$ -- (3/2,1/2)$_2$ and (5/2,1/2)$_3$ -- (5/2,1/2)$_2$
lines differ by only 0.003 nm and thus cannot be resolved. The former of these
two lines has a factor of three higher intensity than the latter, and therefore
it is for this transition that the experimental wavelength is listed in Table
\ref{Table2}. The remaining three lines in the Table are reliably resolved in
the measured spectrum. The [Zn] intercombination $4s^2 - 4s4p$ line at
13.2878 nm is still seen in the spectrum, although this
ion is present at only about 0.3 \% of the total population.

At longer wavelengths, a line at 19.1488~nm was identified as the M1 transition
$3d^94s$ (3/2,1/2)$_1$ -- $3d^94s$ (5/2,1/2)$_2$ in W$^{46+}$. In Ref.
\cite{PRA} we found that the decay of the upper level $3d^94s$ (5/2,1/2)$_2$
through this process  is crucially important for the accurate modeling of
intensities of the $3d^{10}-3d^94s$ forbidden lines at 0.8 nm (see also
\cite{RalJPB}). Of course, this transition also substantially affects the
population of the lower level $3d^94s$ (3/2,1/2)$_1$.

\section{Conclusions}{\label{concl}}

We presented the measurements of the EUV spectra of highly-charged ions of
tungsten in the spectral range from 4 nm to 20 nm, emphasizing the study of
spectral features above 12 nm. Implementing advanced collisional-radiative
modeling based on extensive accurate sets of atomic data, we were able to
identify almost all observed spectral lines. Some of those are newly reported
here. Additional independent studies would be valuable in the case of the
anomalous intensity observed for the Ga-like line at 12.817 nm, the wavelengths
of the poorly resolved Ge-like E1 lines near 13 nm, and the likely previous
misidentification of a Zn-like line at 4.45299 nm \cite{LLNL}.

\ack{\label{ack}}

This work was supported in part by the Office of Fusion Energy Sciences of the
U. S. Department of Energy. We are grateful to A.E. Kramida for valuable
discussions.

\newpage

\begin{landscape}

\begin{table}
\caption{\label{Table2}Measured spectral lines of highly-ionized tungsten
(wavelengths are in nm).
We use notation $a[b]$ for $a \cdot 10^{b}$. The experimental uncertainties for
wavelengths are
given in parentheses.}
\begin{indented}
\lineup
\item[]\begin{tabular}{@{}cccccccccc}
\br
Ion & Transition & Type & \multicolumn{2}{c}{$gf$(Theory)} &
\multicolumn{2}{c}{$\lambda$(Exper.)}
&  \multicolumn{2}{c}{$\lambda$(Theory)} & Note\\
 & & & Ref. \cite{Fournier} & this work & this work & Ref. \cite{put} & Ref. \cite{Fournier}
& this work  & \\
\mr
W$^{39+}$ [Br] & $4p^{5}\ ^{2}P_{3/2} - 4p^{5}\ ^{2}P_{1/2}$ & M1 & 3.94[-5] & 4.12[-5] & 13.474(4) 
& & 13.35922 & 13.347 & new line\\
\\
W$^{40+}$ [Se] &$4p^{4}\ ^{3}P_{2} - 4p^{4}\ ^{1}D_{2}$ & M1 &  4.85[-5] & 5.18[-5] &  12.852(5) 
& 12.864(5) & 12.83020 & 12.847 & \\
 &  $4p^{4}\ ^{3}P_{2} - 4p^{4}\ ^{3}P_{1}$ & M1 &  5.10[-5] & 5.23[-5] & 13.488(9) 
& 13.487(5) & 13.49228 & 13.493 & \\
\\
W$^{41+}$ [As] &  $4p^{3}\ ^{2}D_{3/2} - 4p^{3}\ ^{2}D_{5/2}$ & M1 &  1.90[-5] & 2.28[-5] & 13.124(4) 
& 13.121(5) & 13.23688 & 13.127 & \\
 &  $4p^{3}\ ^{2}D_{3/2} - 4p^{3}\ ^{4}S_{3/2}$ & M1 &  8.16[-5] & 8.23[-5] & 13.914(5) 
& 13.896(5) & 14.10290 & 13.956 & \\
\\
W$^{42+}$ [Ge] &  $4s^{2}4p^{2}\ ^{3}P_{1} - 4s4p^{3}\ ^{3}P_{2}$ & E1 &  1.50[-1] &
1.56[-1] & 12.895(4) 
& 12.912(5) & 12.73583 & 12.768 & tentative \\
 &  $4s^{2}4p^{2}\ ^{3}P_{0} - 4s^{2}4p^{2}\ ^{1}D_{2}$ & E2 &  1.78[-6] & 1.87[-6]
 & 12.941(4) 
& 12.945(5) & 13.05679 & 12.951 & \\
 &  $4s^{2}4p^{2}\ ^{1}D_{2} - 4s4p^{3}\ ^{3}P_{2}$ & E1 &  1.25[-1] & 1.36[-1] &
 13.495(6)
& 13.475(5) & 13.35610 & 13.373 & tentative\\
 &  $4s^{2}4p^{2}\ ^{3}P_{0} - 4s^{2}4p^{2}\ ^{3}P_{1}$ & M1 &  4.09[-5] & 4.09[-5]
 & 13.545(4) 
&  & 13.70953 & 13.574 & 13.475 nm in \cite{put}\\
\\
W$^{43+}$ [Ga] &  $4s^{2}4p\ ^{2}P_{1/2} - 4s^{2}4p\ ^{2}P_{3/2}$ & M1 &  4.17[-5] &
4.37[-5] & 12.629(3) 
& 12.639(5) & 12.60060 & 12.638 & \\
 &  $4s^{2}4p\ ^{2}P_{1/2} - 4s4p^{2}\ ^{4}P_{1/2}$ & E1 &  1.38[-1] & 1.50[-1] & 12.817(4) 
& 12.824(5) & 12.70610 & 12.817 & \\
 &  $4s^{2}4p\ ^{2}P_{3/2} - 4s4p^{2}\ ^{4}P_{5/2}$ & E1 &  2.53[-1] & 2.29[-1]
 & 13.481(10) 
&  & 13.45507 & 13.506 & 13.534 nm in \cite{put}\\
\\
W$^{44+}$ [Zn] &  $4s^{2}\ ^{1}S_{0} - 4s4p\ ^{3}P_{1}$ & E1 &  1.38[-1] & 1.39[-1] &  13.288(3)
& 13.287(5) & 13.20751 & 13.149 & \\
 &  $4s4p\ ^{3}P_{1} - 4s4p\ ^{3}P_{2}$ & M1 &  5.64[-5] & 5.57[-5] &  13.480(3)
&  & 13.36069 & 13.345 & new line\\
\\
W$^{45+}$ [Cu] &  $4s\ ^{2}S_{1/2} - 4p\ ^{2}P_{1/2}$ & E1 &  2.35[-1] & 2.29[-1] &  12.712(3)
& 12.720(5) & 12.62740 & 12.688 & \\
\\
W$^{46+}$ [Ni] &  $3d^{9}4s\ (3/2,1/2)_{1} - 3d^{9}4p\ (3/2,1/2)_{2}$ & E1 & 
2.98[-1] & 2.96[-1] & 12.860(3)
&  & 12.74701 & 12.841 & new line\\
 &  $3d^{9}4s\ (5/2,1/2)_{3} - 3d^{9}4p\ (5/2,1/2)_{2}$ & E1&  4.56[-1] & 4.54[-1] & 
&  & 12.78189 & 12.844 & blended\\
 &  $3d^{9}4s\ (5/2,1/2)_{2} - 3d^{9}4p\ (5/2,1/2)_{3}$ & E1&  4.55[-1] & 4.52[-1] & 12.958(4)
&  & 12.84375 & 12.944 & new line\\
 &  $3d^{9}4s\ (3/2,1/2)_{2} - 3d^{9}4p\ (3/2,1/2)_{2}$ & E1&  2.73[-1] & 2.71[-1] & 13.019(6)
&  & 12.92460 & 13.012 & new line\\
 &  $3d^{9}4s\ (5/2,1/2)_{2} - 3d^{9}4p\ (5/2,1/2)_{2}$ & E1&  1.08[-1] & 1.07[-1] & 13.113(3)
&  & 13.03312 & 13.112 & new line\\
 &  $3d^{9}4s\ (5/2,1/2)_{2} - 3d^{9}4s\ (3/2,1/2)_{1}$ & M1&  3.98[-5] & 3.98[-5] & 19.149(2)
&  & 19.15539 & 19.195 & new line\\
\\
W$^{47+}$ [Co] &  $3d^{9}\ ^{2}D_{3/2} - 3d^{9}\ ^{2}D_{1/2}$ & M1&  5.15[-5] &
3.43[-5] & 18.578(2)
&  & 18.62292 & 18.640 & new line\\

\br
\end{tabular}
\end{indented}
\end{table}

\end{landscape}

\end{document}